\documentclass[epj,english,twocolumn]{webofc}
\usepackage[varg]{txfonts}   
\usepackage{hyperref}        
\usepackage{upgreek}
\woctitle{Heavy Ion Accelerator Symposium 2013}
\begin{document}
\title{AMS measurements of cosmogenic and supernova-ejected
radionuclides in deep-sea sediment cores}
\author{J. Feige\inst{1}\fnsep\thanks{\email{jenny.feige@univie.ac.at}} \and
        A. Wallner\inst{1,2}
	      \and L.K.\ Fifield\inst{2}
	      \and
	G. Korschinek\inst{3} \and
	S. Merchel\inst{4} \and
	G. Rugel\inst{4} \and
	P. Steier\inst{1} \and
	S.R.\ Winkler\inst{1} \and
	R. Golser\inst{1}
}

\institute{University of Vienna, Faculty of Physics, VERA Laboratory, W\"ahringerstrasse 17, 
	1090 Vienna, Austria
\and
        Department of Nuclear Physics, The Australian National University, Canberra, ACT 0200, Australia 
\and
        Physics Department, TU Munich, James-Franck-Stra\ss e, 85748 Garching, Germany 
\and 
	Helmholtz-Zentrum Dresden-Rossendorf, Bautzner Landstra\ss e 400, 01328 Dresden, Germany
          }

\abstract{%
Samples of two deep-sea sediment cores from the Indian Ocean are analyzed with accelerator mass spectrometry (AMS) to search for traces of recent supernova activity $\sim$2~Myr ago. Here, long-lived radionuclides, which are synthesized in massive stars and ejected in supernova explosions, namely $^{26}$Al, $^{53}$Mn and $^{60}$Fe, are extracted from the sediment samples. The cosmogenic isotope $^{10}$Be, which is mainly produced in the Earth’s atmosphere, is analyzed for dating purposes of the marine sediment cores. 
The first AMS measurement results for $^{10}$Be and $^{26}$Al are presented, which represent for the first time a detailed study in the time period of 1.7-3.1~Myr with high time resolution.
Our first results do not support a significant extraterrestrial signal of $^{26}$Al above terrestrial background. However, there is evidence that, like $^{10}$Be, $^{26}$Al might be a valuable isotope for dating of deep-sea sediment cores for the past few million years.

}
\maketitle
\section{Introduction}
\label{intro}
The search for evidence of recent supernova (SN) activity close to the solar system has become a popular matter over
the last years. It was first suggested by Ellis et al.\ (1996) \cite{ellis} 
and Korschinek et al.\ (1996) \cite{gunther96} to
search for geological isotope anomalies. Since then not only terrestrial (e.g. \cite{kniealt}, \cite{fitoussi}, \cite{knie}) 
but also lunar samples
(\cite{cook}, \cite{fimiani}) have been successfully analyzed for traces of $^{60}$Fe, a radionuclide ejected by SN explosions. 
Terrestrial archives with a homogeneous growth rate in an undisturbed environment,
such as ice cores, deep-sea sediments, and ferromanganese crusts and nodules are most suitable to search for SN-produced elements. In those archives  
material is ideally accumulated at a constant rate and a possible SN signal should be fixed to the time of its deposition.
The enhancement of $^{60}$Fe discovered by Knie et al.\ (2004) \cite{knie} in a ferromanganese crust was located in a layer
corresponding to a time range of 1.7-2.6 Myr \cite{ichselber}.
Due to the very low growth rate of 2.37~mm~Myr$^{-1}$ the signal is contained mainly in a layer of 2~mm thickness
and a more precise dating is not possible. The same is true for lunar samples, where no time information is available. The sedimentation rate is negligible and mixing of layers e.g.\ by 
micrometeoritic and dust impacts, the 
so called gardening effect, may disperse any signal \cite{cook}.
Low sedimentation rate is an advantage; the signal is deposited in a very 
thin layer and a small sample volume covers a long time period. If a signature is present, crust samples are better suited for obtaining a first hint on the time of deposition; but there is only a limited time resolution. The search in sediments and ice cores with much higher time resolution, however, requires much more sample material to be processed.
In ice cores it is not possible to date back more than some 100~kyr \cite{epica}. Therefore, these are not suitable archives for a SN trace 2 Myr ago.
Here, we extract radionuclides (namely $^{10}$Be, $^{26}$Al, $^{53}$Mn and $^{60}$Fe)
from deep-sea sediments with accumulation rates higher than ferromanganese
crusts, i.e. in the order of mm~kyr$^{-1}$, to obtain a good time resolution. \par
The findings of an enhancement of SN-ejected radionuclides in a deep-sea sediment would corroborate the theory of a recent SN close to the solar vicinity. Not only the $^{60}$Fe signal identified by \cite{knie} is a valuable hint to such events in the local interstellar medium, but also the existence of the Local Bubble, in which our solar system is embedded \cite{fuchs}. This interstellar cavity has been produced by $\sim$20 SN explosions starting around 14 Myr ago in a stellar moving group of stars belonging today to the the Sco-Cen association. It has been shown, that enough $^{60}$Fe material can be transported from these SN origin to Earth, to produce such a signal as observed in the ferromanganese crust \cite{dieter}.

\section{Procedure}
\label{proc}
\subsection{Sediment Cores and Target Isotopes}
\label{sedis}
Two marine sediment cores from the Indian Ocean, Eltanin piston cores ELT49-53 and ELT45-21, are available for analyzation. These were collected from the South Australian Basin,
38$^\circ$58.7'~S, 103$^\circ$43.1'~E and 37$^\circ$51.6'~S, 100$^\circ$01.7'~E, respectively,
from depths of approximately 4200~m \cite{allled}. The average accumulation rate in these sections was determined to be
2.7~mm~kyr$^{-1}$ for ELT49-53 and 3.7~mm~kyr$^{-1}$ for ELT45-21 by magnetostratigraphic analysis. However, changes in the velocity of bottom water currents triggered by climate change may alter the grain size distribution of the sediment, prevent deposition or even erode deposited material.
Due to these environmental effects the 
sediment accumulation rate varies with time \cite{allled}.
Sections have been provided from depths of 120-697~m below the sea floor, spanning a time interval from 1.7-3.1~Myr. Altogether, there are 26 samples of core ELT45-21 and 45 samples of core ELT49-53 available for analysis. Each sample covers a length of 1~cm, representing an average time period of  $\sim$3~kyr, with distances of 3-17 cm between the individual samples.\par
The radionuclides selected for this work are $^{26}$Al (t$_{1/2}$~=~0.71 Myr) \cite{holden}, $^{53}$Mn (t$_{1/2}$~=~3.7 Myr) \cite{holden}, and $^{60}$Fe (t$_{1/2}$~=~2.6 Myr) \cite{georg}.
All half-lifes are in the order of Myr, and hence all isotopes are potentially useful for exploring the time interval around 2 Myr ago where the peak in $^{60}$Fe concentration was observed in a ferromanganese crust \cite{knie}.\par
Extraterrestrial (i.e.\ stellar) $^{26}$Al is mainly produced in three different environments in massive stars via the $^{25}$Mg(p,$\upgamma$)$^{26}$Al reaction: 
the central core burning in main sequence stars, the C and Ne burning shell in
the later stages of the star, and during explosive Ne burning \cite{limch}.
A natural terrestrial background of $^{26}$Al is constantly generated by spallation
reactions from cosmic rays of Ar in the Earth's atmosphere amounting to a global mean of
1.3$\times$10$^3$ atoms cm$^{-2}$ yr$^{-1}$ \cite{auer}. \par
The dominant stellar production of $^{53}$Mn is within
explosive silicon and oxygen burning inside of 2 M$_\odot$ of a massive star \cite{meyer}. A 
background contribution arises from extraterrestrial input of dust, meteorites and micrometeorites
onto Earth. A target element for the production of $^{53}$Mn by nuclear reactions of 
cosmic ray protons and secondary neutrons is iron, there is also a smaller contribution from nickel \cite{silkeundklaus}.
The annual flux of $^{53}$Mn has been estimated to be approximately 200 atoms cm$^{-2}$ yr$^{-1}$ \cite{auerdiss}.\par
In contrast to $^{26}$Al and $^{53}$Mn, the influx of $^{60}$Fe 
coming from meteorites and micrometeorites should be very low. Here, the target element is nickel only; the production
rate of $^{60}$Fe from cosmic rays is around three orders of magnitude lower than the production rate of 
$^{53}$Mn \cite{silkeundklaus}. 
Stellar $^{60}$Fe is
produced in the late burning stages of massive stars via neutron capture on $^{59}$Fe. The half-life of $^{59}$Fe
is only 44.5 days, hence the production of $^{60}$Fe competes with the beta-decay of $^{59}$Fe.
Hot temperatures above 4$\times$10$^{8}$ K are required to produce a sufficient amount of neutrons limiting the synthesis of $^{60}$Fe to the shell burning of He and C and to explosive Ne burning \cite{limch}.
\par
Estimations of the fluence of these isotopes in the marine sediment cores and the probability of
detecting a potential signal with AMS have been presented earlier \cite{ichselber}. In the case of $^{26}$Al and $^{53}$Mn we would expect an exponential decay profile with increasing depth in a core. A positive variance of the data in the decay curve might indicate a SN signature. The height and broadness of such a
signal depends on various effects such as the amount of material ejected in the SN explosion, the dimension of the incoming shock wave containing the material, the time it takes for the 
radionuclides to be deposited, etc.\par
Additionally to these three radionuclides chosen for SN detection, $^{10}$Be (t$_{1/2}$ = 1.4 Myr) (\cite{guntherhalflife}, \cite{chmeleff})
is measured to obtain a relative dating of both cores. Like cosmogenic $^{26}$Al it is produced in the Earth's
atmosphere. It originates from spallation induced by cosmic rays of mainly nitrogen and oxygen leading to an average total flux of 6.5$\times$10$^5$ atoms cm$^{-2}$ yr$^{-1}$ \cite{masbee}.
\subsection{Sample Preparation}
\label{prep}
Chemical sample preparation was adapted from the leaching procedure described by Bourl\`{e}s et al.\
\cite{bourles} and Fitoussi et al.\ \cite{FitoussiChemie}. This technique is designed to extract the authigenic fraction, which includes the isotopes of interest, from the detrital phase. Dissolving the whole sample material leads to enhanced extraction of the stable elements from the detrital component, which dilute the signal (see section \ref{meas}). Further chemical separation of Al, Be, Fe, and Mn from the leachate is an updated version of
Merchel \& Herpers \cite{silke}, to adjust for higher concentration of elements like Ca, Na, and K originating from a total starting mass of 3~g of sediment.\par
The samples were leached for 1 hour in 60 ml of 0.04~M NH$_2$OH$\cdot$HCl in 25\% (V/V) acetic acid at (90$\pm$5)$^\circ$C,
then for 7 hours at (95$\pm$5)$^\circ$C \cite{bourles}. After taking an aliquot for stable $^9$Be-determination by inductively coupled plasma mass spectrometry (ICP-MS) about 600 $\upmu$g of stable $^9$Be-carrier was added to each leachate. Precipitation with NH$_{3\mathrm{aq}}$ and removal of hydroxides of Al, Be and Fe from the solution resulted in slow oxidation of Mn$^{2+}$ to Mn$^{4+}$ and thus, delayed precipitation and separation of MnO(OH)$_2$ after a few hours. Purification to decrease the isobar ($^{53}$Cr) content is achieved by reprecipitation, i.e. dissolution in HNO$_3$ and H$_2$O$_2$ and precipitation with KClO$_3$ in the heat.
Repeated washing and drying result in targets of MnO$_2$, suitable for AMS measurements. Iron was separated from Al and Be by dissolving in HCl and applying the solution to an anion exchange column \cite{silke}. Nickel containing the isobar of $^{60}$Fe, i.e.\ $^{60}$Ni, which was already mainly reduced in the first precipitation step, is further decreased during this anion exchange.  Al and Be were separated from each other by cation exchange. In this step, the boron-fraction containing $^{10}$B is further reduced. All three elements, Al, Be and Fe, were precipitated with NH$_{3\mathrm{aq}}$ as hydroxides, washed, dried, and for Be and Al ignited to oxides at 900$^\circ$C.
\subsection{AMS Measurements}
\label{meas}
The concentrations of $^{10}$Be and $^{26}$Al $^{53}$Mn and $^{60}$Fe in the marine sediment samples
are measured as isotopic ratios, i.e.\ the ratio of the radionuclide relative to the stable isotope, using accelerator mass spectrometry (AMS).
A beam of negative ions is produced in a Cs sputter source, preaccelerated and analyzed through a combination of an electrostatic analyzer and an injector magnet. Most importantly, isobaric molecular interferences are excluded in AMS by injecting the negative particles into a tandem accelerator, where molecules break up by stripping processes in a foil or gas stripper. The now positively charged high-energy ions pass another mass spectrometer, where break-up products of molecules consisting now of different masses are filtered from the beam (e.g.\ \cite{walter}). Atomic stable isobars, which are much more abundant in nature than the corresponding isobaric radionuclide, will pass through the AMS setup in the same manner, and will be suppressed by specific methods for different elements. Therefore, AMS is highly sensitive and capable of quantifying very small isotopic ratios as low as 10$^{-16}$, which makes it very suitable for a wide range of applications \cite{walterams}, in particular also astrophysical applications \cite{toniastro}. \par
Several different laboratories are involved in the AMS measurements.
$^{26}$Al is analyzed at VERA (Vienna Environmental Research Accelerator), Austria, a 3~MV tandem accelerator \cite{peter}. For AMS Al$^-$ ions are selected. Dedicated isobaric suppression is not necessary, as the stable isobar $^{26}$Mg does not form negative ions.\par
The DREAMS (Dresden AMS) facility \cite{shavkat}, a 6~MV tandem accelerator at Helmholtz-Zentrum Dresden-Rossendorf, Germany, is used for collection of $^{10}$Be/$^{9}$Be data in the marine sediment targets. At this laboratory, $^{10}$B suppression is achieved by a 1 $\upmu$m thick silicon nitride absorber foil placed between the high-energy 90$^\circ$ analyzing magnet and a 35$^\circ$ electrostatic analyzer \cite{shavkat}. Comparative measurements will carried out at VERA. 
\par
Sufficient suppression of the interfering stable isobars $^{53}$Cr and $^{60}$Ni of $^{53}$Mn and $^{60}$Fe, respectively, is only possible by acceleration to very high energies in the order of 100~MeV in combination with a gas-filled magnet (e.g.\ \cite{knie2000}). Separation is achieved by interaction with the gas atoms leading to different average charge states of the radionuclide and its stable isobar depending on their atomic numbers. Therefore, isobars are deflected differently and take separate trajectories in the magnetic field. AMS machines with terminal voltages larger than 10 MV are needed to achieve the energies required for separation. Facilities capable of detecting $^{53}$Mn and $^{60}$Fe at the expected low levels of our study are the 14~MV MP tandem accelerator at the Maier-Leibnitz-Laboratory in Garching, Germany \cite{knie2000} and the HIAF (Heavy Ion Accelerator Facility), a 15~MV pelletron accelerator at the ANU in Canberra, Australia \cite{keith}. $^{53}$Mn and $^{60}$Fe will be measured at the latter laboratory. 
\section{Results}
\label{res}
\begin{figure}[!tb]
\centering
\includegraphics[width=\linewidth]{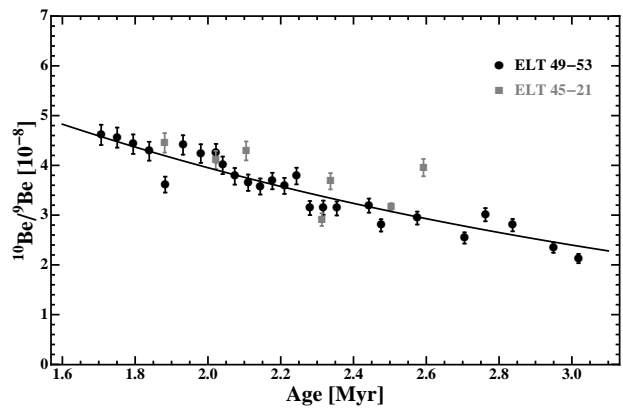}
\caption{The $^{10}$Be/$^{9}$Be data from DREAMS in the two marine sediment cores ELT49-53 (black circles) and ELT45-21 (gray squares) with increasing age, not corrected for half-life. The data fits well to the values expected from exponential decay (solid line).}
\label{be}       
\end{figure}

\begin{figure}[!tb]
\centering
\includegraphics[width=\linewidth]{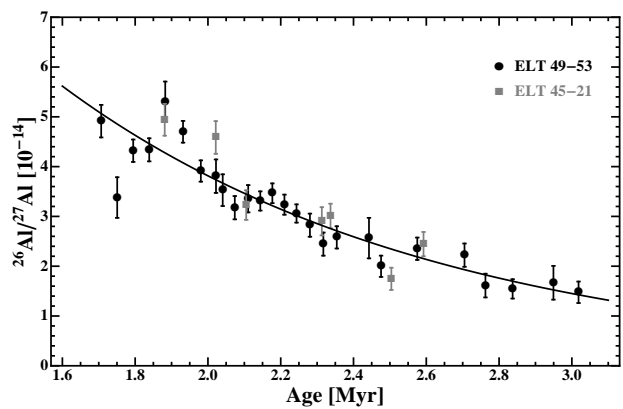}
\caption{The $^{26}$Al/$^{27}$Al data vs age of the two marine sediment cores. Exponential decay is indicated by the solid line.}
\label{al}       
\end{figure}

The distribution of $^{10}$Be/$^{9}$Be in a time range of 1.7-3.1~Myr measured with the DREAMS facility is presented in Fig.~\ref{be}. Here, the in-house standard SMD-Be-12 with a $^{10}$Be/$^{9}$Be value of (1.70$\pm$0.03)$\times$10$^{-12}$ \cite{shavkat} has been used to normalize the data. The error bars originate from a composition of statistical uncertainty from the AMS measurements (which are usually between 1-2 \%) and of the uncertainty from stable isotope ICP-MS measurements. No precise information on the uncertainties of ICP-MS data are available yet; commonly they are in the range of 3-5 \%. First estimations from repeated measurements of several samples lead to an average uncertainty of 4 \%. The preliminary chronology of the sediment cores as indicated in Fig.~\ref{be} and \ref{al} has been extracted from magnetostratigraphic data of \cite{allled}. The $^{10}$Be/$^{9}$Be data of the two cores have been fitted with an exponential decay function using the half-life of $^{10}$Be. Extrapolating this curve to an initial value of $^{10}$Be/$^{9}$Be at the surface results in (1.07$\pm$0.05)$\times$10$^{-7}$. This ratio falls within the range of values presented by \cite{bourles} for the Indian Ocean.
\par
A variability of $^{10}$Be/$^{9}$Be within both cores is indicated in the data shown in Fig.~\ref{be}.
Variations are commonly associated with a change in the production of cosmogenic $^{10}$Be in the Earth's atmosphere, which is induced by the variability of the geomagnetic field (see \cite{didier}).
It is possible, that melting ice after glacial periods releases $^{10}$Be into the ocean, which would lead to an increase of the $^{10}$Be/$^{9}$Be ratio (\cite{aldahan}, \cite{wang}). 
The $^{10}$Be concentration is influenced by climate changes, which might lead to changes in the bottom water circulation and an increase or decrease of the sedimentation rates \cite{allled}. 
It has to be further investigated, which effects play a significant role in our samples.
\par
The $^{26}$Al/$^{27}$Al data for the two sediment cores has been measured with the VERA facility, Vienna. We obtain very low isotopic ratios of $\sim$10$^{-14}$, requiring a long measurement time. Each sample was sputtered for several hours until complete exhaustion to obtain a statistical uncertainty of better than 10~\%. There are several reasons for the small isotopic ratios of $^{26}$Al/$^{27}$Al compared to $^{10}$Be/$^{9}$Be. The atmospheric production rate of $^{26}$Al is approximately 10$^3$ times lower than of $^{10}$Be \cite{auer} and the age of the samples correspond to three half-lifes of $^{26}$Al, which means a lot of $^{26}$Al has already decayed. In addition stable aluminium is a major component in deep-sea clay sediments. This is partially suppressed by leaching rather than dissolving the whole material during the chemical separation. 
\par
The data shown in Fig.~\ref{al} were normalized to the standard material AW-V-2 and AW-V-3 with $^{26}$Al/$^{27}$Al values of (2.71$\pm$0.02)$\times$10$^{-12}$ and (3.65$\pm$0.05)$\times$10$^{-12}$, respectively \cite{toni}. The combined $^{26}$Al/$^{27}$Al distribution throughout the marine sediment cores is fitted with one exponential decay curve based on the half-life of $^{26}$Al. A surface value of (2.6$\pm$0.1)$\times$10$^{-13}$ is deduced. The $^{26}$Al/$^{27}$Al ratios in the two cores seem to agree with each other very well, an indication that $^{26}$Al might be a suitable isotope for dating of marine sediment samples. A significant signal of a recent close-by SN explosion can not be extracted from the presented data.

\section{Conclusions and Outlook}
\label{conc}
We have analyzed two deep-sea sediment cores from the Indian Ocean for a potential SN signal. They span a time period between 1.7 and 3.1 Myr. We studied samples of 3 kyr integration time for every 20-30 kyr.  Radionuclides analyzed so far are $^{26}$Al and $^{10}$Be, where the latter is measured for an independent dating of the individual samples. Our data represents for the first time a detailed study in this time period with high time resolution. $^{53}$Mn and $^{60}$Fe will be measured in the same sediment samples. 
\par
The samples have been chemically processed with a separation technique 
following \cite{bourles} and \cite{silke} to extract the elements Al, Be, Mn, and Fe. To date, most samples have been analyzed by AMS. Measurements of
$^{26}$Al and $^{10}$Be have been performed at the VERA and the DREAMS facilities, respectively. The data are compatible with ratios expected from atmospheric production by cosmic rays. Although more challenging to measure, due to low $^{26}$Al/$^{27}$Al ratios, these adapt to the exponential fit very well. Like $^{10}$Be/$^{9}$Be, the $^{26}$Al/$^{27}$Al data seem to agree with each other in the two sediment cores, making $^{26}$Al a potential radionuclide for dating purposes. A significant SN signal can not be identified yet.
Longer measurements of the $^{26}$Al/$^{27}$Al targets will be carried out to reduce the statistical uncertainty.\par
The $^{10}$Be/$^{9}$Be data presented here was collected with DREAMS. Comparative measurements with VERA are in progress. \par
New sample material of the marine sediments surface will be chemically prepared and analyzed to compare the results with extrapolations from our current measurements and with $^{26}$Al/$^{27}$Al and $^{10}$Be/$^{9}$Be ratios expected in the Indian Ocean. 
A potential SN signal is best identified by measurements of radionuclides with a low background, such as $^{60}$Fe. These measurements are performed at the ANU, Canberra, and will be compared with results from $^{26}$Al and $^{53}$Mn. First measurements of $^{53}$Mn are planned to take place later in 2013, also at ANU, Canberra. 

%
%
\section*{Acknowledgments} 
This work is funded by the Austrian Science Fund (FWF), project P20434 and I428 (EUROCORES project EuroGENESIS, subproject CoDustMas, funded via the European Science Foundation).
This research used samples and data provided by the Antarctic Marine Geology Research Facility (AMGRF) at Florida State University. The AMGRF is sponsored by the U.S. National Science Foundation.
We would like to thank the accelerator staff at 
Dresden-Rossendorf for their support. Furthermore, we thank Aline Ritter (HZDR) for stable isotope measurements and the CEREGE-team, especially Didier Bourl\`{e}s, for sharing detailed information on leaching procedures. 
%

%
\end{document}